\documentclass[
twocolumn,
english,
reprint,
superscriptaddress,
aps,
prd,]
{revtex4-2}

\usepackage[T1]{fontenc}
\usepackage[latin9]{inputenc}
\usepackage{color}
\usepackage{babel}
\usepackage{textcomp}
\usepackage{mathrsfs}
\usepackage{dsfont}
\usepackage{amsmath}
\usepackage{amssymb}
\usepackage{graphicx}
\usepackage[unicode=true,pdfusetitle,
 bookmarks=true,bookmarksnumbered=false,bookmarksopen=false,
 breaklinks=false,pdfborder={0 0 0},pdfborderstyle={},backref=false,colorlinks=true]
 {hyperref}

\makeatletter

\newcommand{\lyxmathsym}[1]{\ifmmode\begingroup\def\b@ld{bold}
  \text{\ifx\math@version\b@ld\bfseries\fi#1}\endgroup\else#1\fi}

\providecommand{\tabularnewline}{\\}

\makeatother

\begin{document}
\title{Bell nonlocality and entanglement in $e^{+}e^{-} \rightarrow Y\bar{Y}$
at BESIII}

\author{Sihao Wu}
\email{shwu@mail.ustc.edu.cn}

\affiliation{Department of Modern Physics and Anhui Center for fundamental sciences
in theoretical physics, University of Science and Technology of China,
Hefei 230026, China}
\author{Chen Qian}
\email{qianchen@baqis.ac.cn}

\affiliation{Beijing Academy of Quantum Information Sciences, Beijing 100193, China}
\author{Qun Wang}
\email{qunwang@ustc.edu.cn}

\affiliation{Department of Modern Physics and Anhui Center for fundamental sciences
in theoretical physics, University of Science and Technology of China,
Hefei 230026, China}
\affiliation{School of Mechanics and Physics, Anhui University of Science and Technology,
Huainan,Anhui 232001, China}
\author{Xiao-Rong Zhou}
\email{zxrong@ustc.edu.cn}

\affiliation{Department of Modern Physics, University of Science and Technology
of China, Hefei 230026, China}

\date{\today}

\begin{abstract}
The Bell nonlocality and entanglement are two kinds of quantum correlations
in quantum systems. Due to the recent upgrade in Beijing Spectrometer
III (BESIII) experiment, it is possible to explore the nonlocality
and entanglement in hyperon-antihyperon systems produced in electron-positron
annihilation with high precision data. We provide a systematic method
for studying quantum correlations in spin-1/2 hyperon-antihyperon
systems through the measures for the nonlocality and entanglement.
We find that with nonvanishing polarizations of the hyperon and its
antihyperon, the kinematic region of nonlocality in the hyperon-antihyperon
system is more restricted than the $\tau^{+}\tau^{-}$ system in which
polarizations of $\tau$ leptons are vanishing. We also present an
experimental proposal to probe the nonlocality and entanglement in
hyperon-antihyperon systems at BSEIII.
\end{abstract}

\maketitle


\section{Introduction}

Quantum mechanics, as a foundational pillar for modern physics, governs
the properties of fundamental particles and their interactions. In
this context, quantum information properties of fundamental particles
can offer a novel perspective on understanding quantum mechanics.
The Bell nonlocality, characterized by the violation of Bell-type
inequalities~\citep{Bell:1964kc,Clauser:1974tg,Aspect:1981zz}, is
a distinctive quantum property with significant implications for quantum
mechanics. Closely related to the Bell nonlocality, the quantum entanglement
is an invisible link between two particles that allows one to instantly
affect the other regardless of their distance. The entanglement has
practical applications in quantum information processing, including
quantum computing~\citep{Jozsa:2003proceeding}, quantum metrology~\citep{Giovannetti:2011naturephotonics},
and quantum communication~\citep{Curty:2004prl}. In the research
area of quantum information theory, theoretical details of the Bell
nonlocality and entanglement have been thoroughly discussed (see,
e.g., Refs.~\citep{Brunner:2014rmp,Horodecki:2009rmp} for recent
reviews). Historically, the Bell nonlocality and entanglement have
been widely studied in photonic and atomic systems~\citep{Yin:2017ips,BIGBellTest:2018ebd}.


High-energy colliders provide an alternative testing ground for the
nonlocality and entanglement~\citep{Aspect2002}. The significant
improvement in collider and detector technology has led to a large
collection of high precision data, thereby enabling the possibility
of observing the quantum correlation in high energy processes. Recently
quantum correlations in elementary particle systems, e.g., top quark
pairs at Large Hadron Collider (LHC)~\citep{Aoude:2022imd,Afik:2020onf,Afik:2022dgh,Afik:2022kwm,Fabbrichesi:2021npl},
leptons pairs~\citep{Fabbrichesi:2022ovb,Ehataht:2023zzt}, gauge
bosons from Higgs decay~\citep{Barr:2021zcp,Barr:2022wyq,Aguilar-Saavedra:2022wam,Fabbrichesi:2023cev},
have been investigated.


In contrast to elementary particles, the use of hadronic final states
to test quantum correlations has a relatively long history, dating
back to early 1980s~\citep{Tornqvist:1980af}. Subsequent studies
came up in the past decades aiming at probing quantum correlations
in hyperon systems~\citep{Tornqvist:1986vv,Baranov:2009zza,Chen:2013epa,Shi:2019kjf,Qian:2020ini,Gong:2021bcp,Lv:2024uev}.
The hyperon's weak decay can serve as its own polarimeter and make
it possible to extract spin observable in the hyperon-antihyperon
system, including polarization and correlation, in experiments. With
the recent upgrade of Beijing Spectrometer III (BESIII) at Beijing
electron-positron collider, there is considerable potential to explore
quantum correlations in hyperon-antihyperon production processes in
electron-positron annihilation~\citep{BESIII:2018cnd,BESIII:2021ypr,Schonning:2023mge}.


In this paper, we investigate the Bell nonlocality and entanglement
in $e^{+}e^{-}\rightarrow\gamma^{*}/\psi\rightarrow Y\bar{Y}$ processes
at BESIII, where $Y$ and $\bar{Y}$ denote the spin-1/2 hyperon and
its antihyperon respectively. Our study is based on the two-qubit
density operator~\citep{Perotti:2018wxm,Batozskaya:2023rek} for
$Y\bar{Y}$. Unlike elementary particle systems such as $\tau^{+}\tau^{-}$
at Belle II and $t\bar{t}$ at LHC, the existence of electromagnetic
form factors (EMFFs) in a polarized $Y\bar{Y}$ state at BESIII~\citep{Faldt:2017kgy}
makes the $Y\bar{Y}$ system different from elementary particle systems~\citep{Ehataht:2023zzt,Afik:2022kwm}.
Recognizing that the final $Y\bar{Y}$ state is local unitary equivalent
to the two-qubit $X$ state, we will derive the analytical expressions
of nonlocality and entanglement for $Y\bar{Y}$. At the end of this
paper, we will discuss the effect of EMFFs in quantum correlation
and also give a proposal to probe the nonlocality and entanglement
at BESIII.


This paper is organized as follows. We will introduce the two-qubit
density operator for $Y\bar{Y}$ produced in electron-positron annihilation
in Sec.~\ref{sec:Preliminaries}. In Sec.~\ref{sec:Nonlocality},
we will discuss the two-qubit $X$ state and investigate the Bell
nonlocality for $Y\bar{Y}$. The quantum entanglement in $Y\bar{Y}$
will be addressed in Sec.~\ref{sec:Entanglement}. The relation between
the Bell nonlocality and entanglement will be discussed in Sec.~\ref{sec:Discussions}.
In Sec. \ref{sec:experiment}, we will give a proposal to probe the
nonlocality and entanglement at BESIII. The final section, Sec.~\ref{sec:Summary},
presents a summary of main results and outlook for future directions
of study.


\section{Preliminaries \label{sec:Preliminaries}}

Hyperon-antihyperon pairs can be produced in electron-positron annihilation
either through the virtual photon exchange $e^{+}e^{-}\rightarrow\gamma^{*}\rightarrow Y\bar{Y}$
or through vector charmonium decays, e.g., $e^{+}e^{-}\rightarrow J/\psi\rightarrow Y\bar{Y}$,
where $Y$ denotes a ground-state octet hyperon $\Lambda$, $\Sigma^{+}$,
$\Xi^{-}$ or \textbf{$\Xi^{0}$}. In BESIII experiments, a huge number
of events for vector charmonia $J/\psi$ and $\psi(2S)$ have been
collected. These vector charmonia can decay into hyperon-antihyperon
pairs. A $Y\bar{Y}$ pair made of two spin-1/2 particles forms a massive
two-qubit system. Due to momentum conservation, in the center of mass
(CM) frame, the outgoing hyperon and antihyperon are back-to-back
in momentum. Their spin states can be characterized by a two-qubit
density operator 
\begin{align}
\rho_{Y\bar{Y}}= & \frac{1}{4}\biggl(\mathds{1}\otimes\mathds{1}+\mathbf{P}^{+}\cdot\boldsymbol{\sigma}\otimes\mathds{1}+\mathds{1}\otimes\mathbf{P}^{-}\cdot\boldsymbol{\sigma}\nonumber \\
 & +\sum_{i,j}C_{ij}\sigma_{i}\otimes\sigma_{j}\biggr),\label{eq:two_qubit}
\end{align}
with $\boldsymbol{\sigma}=(\sigma_{1},\sigma_{2},\sigma_{3})$ being
Pauli matrices, $\mathbf{P}^{\pm}$ the polarization or Bloch vectors
of hyperon/antihyperon, and $C_{ij}$ their correlation matrix. The
two-qubit density operator Eq.~(\ref{eq:two_qubit}) can also be put into
a more compact form: $\rho_{Y\bar{Y}}=(1/4)\Theta_{\mu\nu}\sigma_{\mu}\otimes\sigma_{\nu}$
with $\Theta_{00}=1$, $\Theta_{i0}=P_{i}^{+}$, $\Theta_{0j}=P_{j}^{-}$,
and $\Theta_{ij}=C_{ij}$. Here, $\sigma_{0}$ is defined as the $2\times2$
identity matrix $\mathds{1}$. In $\rho_{Y\bar{Y}}$ there are $15$
real parameters for the spin configuration of the $Y\bar{Y}$ pair. 

\begin{figure}[ht]
\includegraphics[scale=0.25]{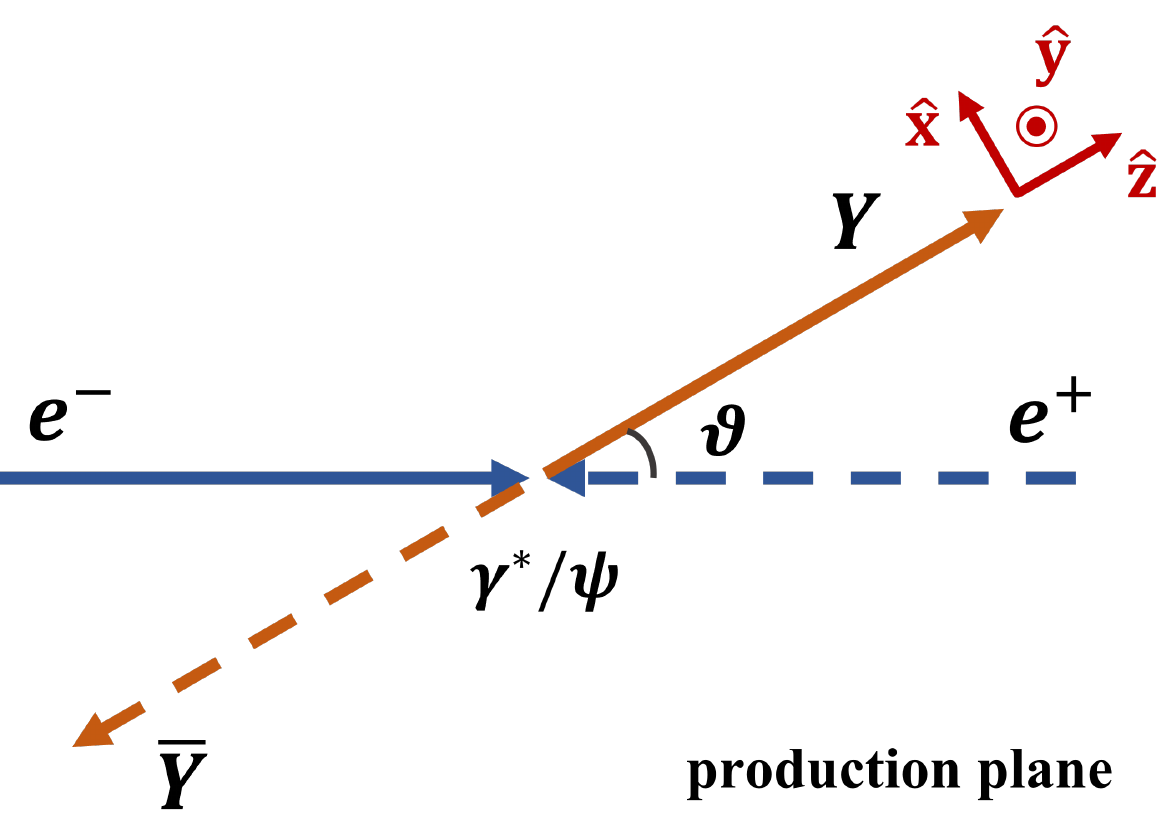}

\caption{\label{fig:frame}The coordinate system used in the analysis with
$\{\hat{\mathbf{x}},\hat{\mathbf{y}},\hat{\mathbf{z}}\}$ being three
directions in the rest frame of $Y$ as well as that of $\bar{Y}$.}
\end{figure}

The $4\times4$ matrix $\Theta_{\mu\nu}$ is frame-dependent. For
the hyperon $Y$, we choose its helicity rest frame as
\begin{equation}
\hat{\mathbf{y}}=\frac{\hat{\mathbf{p}}_{e}\times\hat{\mathbf{p}}_{Y}}{\left|\hat{\mathbf{p}}_{e}\times\hat{\mathbf{p}}_{Y}\right|},\ \hat{\mathbf{z}}=\hat{\mathbf{p}}_{Y},\ \hat{\mathbf{x}}=\hat{\mathbf{y}}\times\hat{\mathbf{z}},\label{eq:helicity-rest-frame}
\end{equation}
which is shown in Fig.~\ref{fig:frame}. While for the antihyperon
$\bar{Y}$, we also adopt its rest frame, but three axes are chosen
to be the same as the hyperon's: $\{\hat{\mathbf{x}}_{\bar{Y}},\hat{\mathbf{y}}_{\bar{Y}},\hat{\mathbf{z}}_{\bar{Y}}\}=\{\hat{\mathbf{x}},\hat{\mathbf{y}},\hat{\mathbf{z}}\}$.
The three axes we choose are different from Refs.~\citep{Perotti:2018wxm,Batozskaya:2023rek},
resulting in slightly different entries of $\Theta_{\mu\nu}$. Adopting
this coordinate system is convenient since the rest frames of $Y$
and $\bar{Y}$ differ only by a pure boost along their momenta without
rotation.

\begin{widetext}
In the rest frames of $Y$ and $\bar{Y}$ with three axes in Eq.~(\ref{eq:helicity-rest-frame}),
through virtual photon exchange $\Theta_{\mu\nu}$ has the form \citep{Perotti:2018wxm,Batozskaya:2023rek}
\begin{equation}
\Theta_{\mu\nu}=\frac{1}{1+\alpha_{\psi}\cos^{2}\vartheta}\left[\begin{array}{c|ccc}
1+\alpha_{\psi}\cos^{2}\vartheta & 0 & \beta_{\psi}\sin\vartheta\cos\vartheta & 0\\
\hline 0 & \sin^{2}\vartheta & 0 & \gamma_{\psi}\sin\vartheta\cos\vartheta\\
\beta_{\psi}\sin\vartheta\cos\vartheta & 0 & -\alpha_{\psi}\sin^{2}\vartheta & 0\\
0 & \gamma_{\psi}\sin\vartheta\cos\vartheta & 0 & \alpha_{\psi}+\cos^{2}\vartheta
\end{array}\right],\label{eq:4by4}
\end{equation}
where $\vartheta$ is the angle between the incoming electron's and
outgoing hyperon's momenta with $\cos\vartheta=\hat{\mathbf{p}}_{e}\cdot\hat{\mathbf{p}}_{Y}$.
Here $\hat{\mathbf{p}}_{e}$ and $\hat{\mathbf{p}}_{Y}$ are momentum
directions of the electron and hyperon respectively. In Eq.~(\ref{eq:4by4}),
$\alpha_{\psi}\in[-1,+1]$ is the decay parameter of the vector charmonium
$\psi(c\bar{c})$, and $\beta_{\psi}$ and $\gamma_{\psi}$ are defined
as 
\begin{equation}
\beta_{\psi}=\sqrt{1-\alpha_{\psi}^{2}}\sin(\Delta\Phi),\ \gamma_{\psi}=\sqrt{1-\alpha_{\psi}^{2}}\cos(\Delta\Phi),
\end{equation}
where $\Delta\Phi\in(-\pi,+\pi]$ is the relative form factor phase. 
\end{widetext}


The polarization and correlation can be read out from $\Theta_{\mu\nu}$
in Eq.~(\ref{eq:4by4}) 
\begin{align}
P_{y}^{+} & =P_{y}^{-}=\frac{\beta_{\psi}\sin\vartheta\cos\vartheta}{1+\alpha_{\psi}\cos^{2}\vartheta},\label{eq:py}
\end{align}
and 
\begin{align}
C_{xx} & =\frac{\sin^{2}\vartheta}{1+\alpha_{\psi}\cos^{2}\vartheta},\ C_{yy}=\frac{-\alpha_{\psi}\sin^{2}\vartheta}{1+\alpha_{\psi}\cos^{2}\vartheta},\nonumber \\
C_{zz} & =\frac{\alpha_{\psi}+\cos^{2}\vartheta}{1+\alpha_{\psi}\cos^{2}\vartheta},\nonumber \\
C_{xz} & =C_{zx}=\frac{\sqrt{1-\alpha_{\psi}^{2}}\cos(\Delta\Phi)\sin\vartheta\cos\vartheta}{1+\alpha_{\psi}\cos^{2}\vartheta}.
\end{align}
Here $P_{y}^{+}$ and $P_{y}^{-}$ are the polarization of $\bar{Y}$
and $\bar{Y}$ along the direction $\hat{\mathbf{y}}$ (the normal
direction of the production plane), respectively. The symmetry property
of the polarization and correlation arises from the invariance under
parity transformation and charge conjugation. We do not consider $\mathcal{CP}$
violation in our analysis.


\section{Bell nonlocality \label{sec:Nonlocality}}

In this section, we will use the hyperon-antihyperon spin density
operator to investigate Bell nonlocality in the $Y\bar{Y}$ system.

\subsection{Local unitary equivalence and $X$ states}

Before our investigation of Bell nonlocality, it is convenient to
transform the two-qubit state in Eqs.~(\ref{eq:two_qubit}) and (\ref{eq:4by4})
to the $X$ state. First, we swap the $\hat{\mathbf{y}}$ and $\hat{\mathbf{z}}$
axes in $Y$ and $\bar{Y}$'s rest frame. Then we diagonalize $C_{ij}$
for $Y$ and $\bar{Y}$. The transformed spin density operator can
be written in terms of Pauli matrices as 
\begin{align}
\rho_{Y\bar{Y}}^{X}= & \frac{1}{4}\biggl(\mathds{1}\otimes\mathds{1}+a\sigma_{z}\otimes\mathds{1}+\mathds{1}\otimes a\sigma_{z}\nonumber \\
 & +\sum_{i}t_{i}\sigma_{i}\otimes\sigma_{i}\biggr),\label{eq:rho_X}
\end{align}
which is in the standard form of a symmetric \textit{two-qubit $X$
state}~\citep{Ting:2007qic}. Thus we place the superscript ``$X$''
to $\rho_{Y\bar{Y}}$. The corresponding $\Theta_{\mu\nu}$ becomes
\begin{equation}
\Theta_{\mu\nu}^{X}=\left[\begin{array}{c|ccc}
1 & 0 & 0 & a\\
\hline 0 & t_{1} & 0 & 0\\
0 & 0 & t_{2} & 0\\
a & 0 & 0 & t_{3}
\end{array}\right],\label{eq:Theta_X}
\end{equation}
where the elements $a$ and $t_{i}$ ($i=1,2,3$) are given by 
\begin{align}
a & =\frac{\beta_{\psi}\sin\vartheta\cos\vartheta}{1+\alpha_{\psi}\cos^{2}\vartheta},\nonumber \\
t_{1,2} & =\frac{1+\alpha_{\psi}\pm\sqrt{\left(1+\alpha_{\psi}\cos2\vartheta\right)^{2}-\beta_{\psi}^{2}\sin^{2}2\vartheta}}{2(1+\alpha_{\psi}\cos^{2}\vartheta)},\nonumber \\
t_{3} & =\frac{-\alpha_{\psi}\sin^{2}\vartheta}{1+\alpha_{\psi}\cos^{2}\vartheta}.\label{eq:a_and_t}
\end{align}
We note that $a=P_{y}^{\pm}$ , $t_{3}=C_{yy}$, and $t_{1,2}$ come
from diagonalizing the block matrix of $C_{ij}$ with $i,j=x,z$ in
$\Theta_{\mu\nu}$.


We note that the swapping of $\hat{\mathbf{y}}$ and $\hat{\mathbf{z}}$
axes and diagonalizing $C_{ij}$ can be obtained by a local unitary
transformation: 
\begin{equation}
\rho_{Y\bar{Y}}^{X}=(U_{Y}\otimes U_{\bar{Y}})\rho_{Y\bar{Y}}(U_{Y}\otimes U_{\bar{Y}})^{\dagger},
\end{equation}
where $U_{Y}$ and $U_{\bar{Y}}$ are two unitary operators acting
independently in $Y$ and $\bar{Y}$'s Hilbert space respectively~\citep{Horodecki:1996pra}.
The states described by $\rho_{Y\bar{Y}}$ and $\rho_{Y\bar{Y}}^{X}$
are said to be \textit{local unitary equivalent} in the sense that
they have same quantum correlation properties such as Bell nonlocality
and entanglement~\citep{Dur:2000pra}. In the remainder of this paper,
all analyses are based on the $X$ state in Eqs.~(\ref{eq:Theta_X})
and (\ref{eq:rho_X}).


\subsection{Bell nonlocality}

The nonlocal property in a quantum entangled system can be tested
by the violation of Bell inequality~\citep{Bell:1964kc}. The most
widely used Bell-type inequality is the CHSH inequality~\citep{Clauser:1969prl}
\begin{equation}
\left|\left\langle A_{1}\otimes B_{1}\right\rangle +\left\langle A_{1}\otimes B_{2}\right\rangle +\left\langle A_{2}\otimes B_{1}\right\rangle -\left\langle A_{2}\otimes B_{2}\right\rangle \right|\leq2,
\end{equation}
where $A_{i}=\mathbf{a}_{i}\cdot\boldsymbol{\sigma}$, $B_{i}=\mathbf{b}_{i}\cdot\boldsymbol{\sigma}$,
and $\langle A_{i}\otimes B_{j}\rangle\equiv\mathrm{Tr}\left[\rho(\mathbf{a}_{i}\cdot\boldsymbol{\sigma}\otimes\mathbf{b}_{j}\cdot\boldsymbol{\sigma})\right]$
with $i,j=1,2$. Here $\mathbf{a}_{1}$, $\mathbf{a}_{2}$, $\mathbf{b}_{1}$
and $\mathbf{b}_{2}$ are four directions (unit vectors) along which
the spin polarization is measured. Then the inequality can be rewritten
in a simpler form 
\begin{equation}
\left|\mathbf{a}_{1}^{\mathrm{T}}C\left(\mathbf{b}_{1}+\mathbf{b}_{2}\right)+\mathbf{a}_{2}^{\mathrm{T}}C\left(\mathbf{b}_{1}-\mathbf{b}_{2}\right)\right|\leq2,\label{eq:Bell_ineq}
\end{equation}
with $C$ being the correlation matrix $C_{ij}$ in Eq.~(\ref{eq:two_qubit}).
Those quantum states that violate the CHSH inequality are called \textit{Bell
nonlocal} states. The maximum of the left-hand side of Eq\@.~(\ref{eq:Bell_ineq})
can be obtained by tuning $\mathbf{a}_{1}$, $\mathbf{a}_{2}$, $\mathbf{b}_{1}$
and $\mathbf{b}_{2}$ as 
\begin{align}
\mathcal{B}\left[\rho\right] & \equiv\max_{\mathbf{a}_{1},\mathbf{a}_{2},\mathbf{b}_{1},\mathbf{b}_{2}}\left|\mathbf{a}_{1}^{\mathrm{T}}C\left(\mathbf{b}_{1}+\mathbf{b}_{2}\right)+\mathbf{a}_{2}^{\mathrm{T}}C\left(\mathbf{b}_{1}-\mathbf{b}_{2}\right)\right|\nonumber \\
 & =2\sqrt{m_{1}+m_{2}},\label{eq:Bell_violation}
\end{align}
where $m_{1}$ and $m_{2}$ are two largest eigenvalues of $C^{\mathrm{T}}C$~\citep{Horodecki:1995nsk}.
Therefore, the CHSH inequality can be violated iff (if and only if)
$m_{1}+m_{2}>1$, and the maximum possible violation of the CHSH inequality
is the upper bound value $2\sqrt{2}$. For convenience, we define
a function of two-qubit density operator $\mathfrak{m}_{12}[\rho]\equiv m_{1}+m_{2}\in[0,2]$
to be a measure of the Bell nonlocality~\citep{Fabbrichesi:2021npl,Ehataht:2023zzt}.


Since we have put the density operator into the $X$ form in (\ref{eq:Theta_X}),
the correlation matrix is diagonal: $\mathbf{t}=\mathrm{diag}\{t_{1},t_{2},t_{3}\}$.
The three eigenvalues of $C^{\mathrm{T}}C$ or $\mathbf{t}^{\mathrm{T}}\mathbf{t}$
are $t_{1}^{2}$, $t_{2}^{2}$ and $t_{3}^{2}$. Then, according to
Eq.~(\ref{eq:Bell_violation}), one needs to select the largest two
values among them. 

\begin{table}
\caption{\label{tab:decay_parameters}Some parameters in $e^{+}e^{-}\rightarrow J/\psi\rightarrow Y\bar{Y}$,
where $Y\bar{Y}$ is a pair of ground-state octet hyperons.}

\begin{ruledtabular}
\begin{tabular}{ccccc}
 & $\mathscr{B}(\times10^{-4})$ & $\alpha_{\psi}$ & $\Delta\Phi/\mathrm{rad}$ & Ref\tabularnewline
\hline 
$\Lambda\bar{\Lambda}$ & $19.43(33)$ & $0.475(4)$ & $0.752(8)$ & \citep{BESIII:2018cnd,BESIII:2017kqw}\tabularnewline
$\Sigma^{+}\bar{\Sigma}^{-}$ & $15.0(24)$ & $-0.508(7)$ & $-0.270(15)$ & \citep{BES:2008hwe,BESIII:2020fqg}\tabularnewline
$\Xi^{-}\bar{\Xi}^{+}$ & $9.7(8)$ & $0.586(16)$ & $1.213(49)$ & \citep{BESIII:2021ypr,ParticleDataGroup:2022pth}\tabularnewline
$\Xi^{0}\bar{\Xi}^{0}$ & $11.65(4)$ & $0.514(16)$ & $1.168(26)$ & \citep{BESIII:2016nix,BESIII:2023drj}\tabularnewline
\end{tabular}
\end{ruledtabular}

\end{table}


As we can see from Eq.~(\ref{eq:a_and_t}) that $t_{1,2,3}$ are
functions of three parameters $\alpha_{\psi}$, $\Delta\Phi$ and
$\vartheta$. Since $t_{1}^{2}\geq t_{2}^{2}$ always holds for any
values of $\alpha_{\psi}$, $\Delta\Phi$ and $\vartheta$, $t_{1}^{2}$
should note be the smallest one. Then, one needs to compare $t_{2}^{2}$
and $t_{3}^{2}$. If $\alpha_{\psi}\geq0$, we aways have $t_{2}^{2}\geq t_{3}^{2}$.
Therefore, the measure of nonlocality becomes $\mathfrak{m}_{12}[\rho_{Y\bar{Y}}^{X}]=t_{1}^{2}+t_{2}^{2}$.
However, for $\alpha_{\psi}<0$, one can not judge which is larger
$t_{2}^{2}$ or $t_{3}^{2}$, since it depends on the specific values
of three parameters. In this case the measure of nonlocality can be
expressed as $\mathfrak{m}_{12}[\rho_{Y\bar{Y}}^{X}]=\max\left\{ t_{1}^{2}+t_{2}^{2},t_{1}^{2}+t_{3}^{2}\right\} $.
In summary, the measure of the Bell nonlocality reads 
\begin{equation}
\mathfrak{m}_{12}\left[\rho_{Y\bar{Y}}^{X}\right]=\begin{cases}
t_{1}^{2}+t_{2}^{2}, & \alpha_{\psi}\geq0\\
\max\left\{ t_{1}^{2}+t_{2}^{2},t_{1}^{2}+t_{3}^{2}\right\} , & \alpha_{\psi}<0
\end{cases}\label{eq:m_12}
\end{equation}
where $t_{1}^{2}+t_{2}^{2}$ and $t_{1}^{2}+t_{3}^{2}$ are given
by 
\begin{equation}
t_{1}^{2}+t_{2}^{2}=1+\left(\frac{\alpha_{\psi}\sin^{2}\vartheta}{1+\alpha_{\psi}\cos^{2}\vartheta}\right)^{2}-2\left(\frac{\beta_{\psi}\sin\vartheta\cos\vartheta}{1+\alpha_{\psi}\cos^{2}\vartheta}\right)^{2},\label{eq:t_12}
\end{equation}
\begin{align}
t_{1}^{2}+t_{3}^{2}= & \left(\frac{1+\alpha_{\psi}+\sqrt{(1+\alpha_{\psi}\cos2\vartheta)^{2}-\beta_{\psi}^{2}\sin^{2}2\vartheta}}{2(1+\alpha_{\psi}\cos^{2}\vartheta)}\right)^{2}\nonumber \\
 & +\frac{\alpha_{\psi}^{2}(1-\cos2\vartheta)^{2}}{4(1+\alpha_{\psi}\cos^{2}\vartheta)^{2}}.
\end{align}

\begin{figure}[ht]
\includegraphics[scale=0.46]{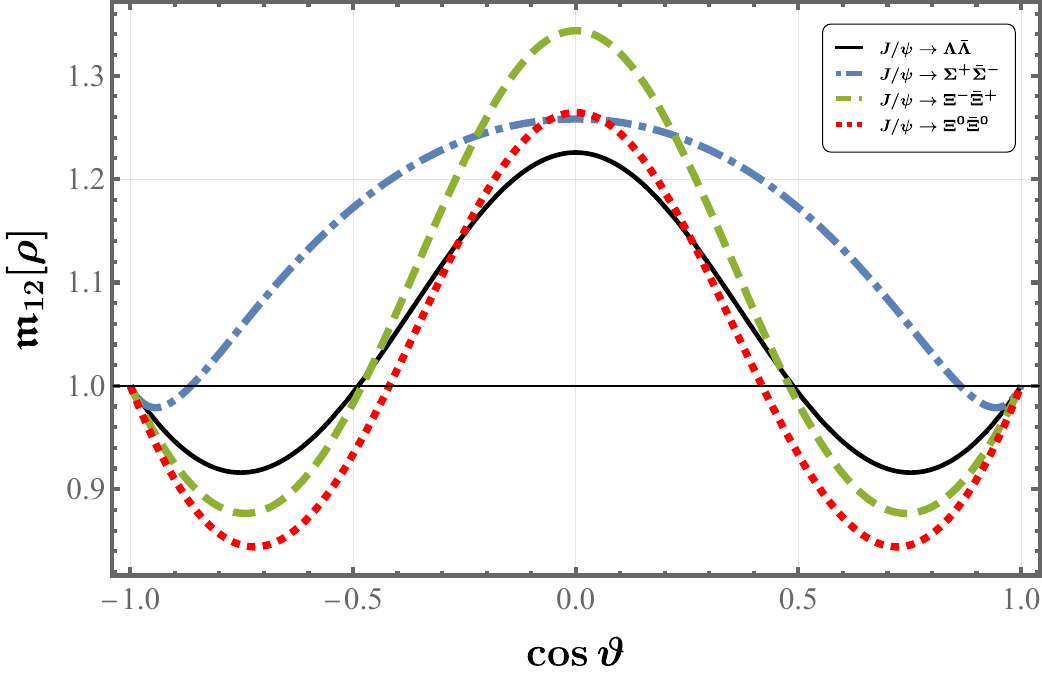}

\caption{\label{fig:m_12}The measure of the Bell nonlocality $\mathfrak{m}_{12}[\rho_{Y\bar{Y}}^{X}]$
as functions of $\cos\vartheta$ ($\vartheta$ is the scattering angle)
in $e^{+}e^{-}\rightarrow J/\psi\rightarrow Y\bar{Y}$ with $Y=\Lambda$,
$\Sigma^{+}$, $\Xi^{-}$ and $\Xi^{0}$ corresponding to curves in
black solid, blue dash-dotted, green dashed, and red dotted lines
respectively. The black horizontal line is the nonlocality bound $\mathfrak{m}_{12}=1$.
The CHSH inequality is violated iff $\mathfrak{m}_{12}>1$.}
\end{figure}


In Table~\ref{tab:decay_parameters} are listed the values of $\alpha_{\psi}$
and $\Delta\Phi$ for $J/\psi$'s decays into a pair of octet hyperons
in electron-positron annihilation. According to these parameters,
we plot $\mathfrak{m}_{12}$ as a function of the scattering angle
$\vartheta$ in Fig.~\ref{fig:m_12} for different decay channels.
From Fig.~\ref{fig:m_12}, we find that $\mathfrak{m}_{12}$ is a
symmetric function of $\vartheta$ relative to $\vartheta=\pi/2$
in the range $\vartheta\in[0,\pi]$, and it reaches the maximum value
$1+\alpha_{\psi}^{2}$ at $\vartheta=\pi/2$. Thus we obtain 
\begin{equation}
\max_{\vartheta}\,\mathcal{B}\left[\rho_{Y\bar{Y}}^{X}\right]=2\sqrt{1+\alpha_{\psi}^{2}}.\label{eq:max_violation}
\end{equation}
By solving $\mathfrak{m}_{12}>1$ in Eq.~(\ref{eq:m_12}) with fixed
$\alpha_{\psi}$ and $\Delta\Phi$, we obtain the nonlocality range
of the scattering angle as $\left(\vartheta^{*},\pi-\vartheta^{*}\right)$.
For $\alpha_{\psi}\geq0$, we can have an analytical expression for
the critical angle $\vartheta^{*}$
\begin{equation}
\vartheta^{*}=\arctan\left|\sqrt{2-2\alpha_{\psi}^{2}}\frac{\sin\Delta\Phi}{\alpha_{\psi}}\right|,\ \ \mathrm{for}\ \alpha_{\psi}\geq0.
\end{equation}
The maximum violation in Eq.~(\ref{eq:max_violation}) and critical
angles in different decay channels are listed in Table~\ref{tab:critical_angle}.

\begin{table}
\caption{\label{tab:critical_angle}The maximum violation $\mathcal{B}_{\mathrm{max}}$ in Eq.~(\ref{eq:max_violation}) and critical angles $\vartheta^{*}$ for the CHSH inequality 
in $e^{+}e^{-}\rightarrow J/\psi\rightarrow Y\bar{Y}$.} 
\begin{ruledtabular}
\begin{tabular}{ccccc}
 & $\Lambda\bar{\Lambda}$ & $\Sigma^{+}\bar{\Sigma}^{-}$ & $\Xi^{-}\bar{\Xi}^{+}$ & $\Xi^{0}\bar{\Xi}^{0}$\tabularnewline
\hline 
$\mathcal{B}_{\mathrm{max}}$ & 2.214 & 2.243 & 2.318 & 2.249\tabularnewline
$\vartheta^{*}$ & 60.81\textdegree & 30.29\textdegree & 61.37\textdegree & 65.27\textdegree
\tabularnewline
\end{tabular}
\end{ruledtabular}
\end{table}


\section{Quantum entanglement \label{sec:Entanglement}}

In this section we will discuss the quantum entanglement in the $Y\bar{Y}$
system and its relation to the Bell nonlocality.

\subsection{Entanglement measure and concurrence }

For a bipartite quantum system living in the combined Hilbert space
$\rho_{AB}\in\mathscr{H}_{A}\otimes\mathscr{H}_{B}$, the state is
said to be \textit{separable} iff the following decomposition holds
\begin{equation}
\rho_{AB}=\sum_{k}p_{k}\rho_{A}^{k}\otimes\rho_{B}^{k},
\end{equation}
where $p_{k}\geq0$ and $\sum_{k}p_{k}=1$, and $\rho_{A}^{k}$ and
$\rho_{B}^{k}$ are the density operator of the corresponding subsystem
$A$ and $B$, respectively. Moreover, the state cannot be decomposed
into the above form is called \textit{non-separable} or \textit{entangled}.


For two-qubit and qubit-qutrit systems ($2\times2$ and $2\times3$
respectively), the Peres-Horodecki criterion provides a sufficient
and necessary condition for separability~\citep{Peres:1996prl,Horodecki:1996pla}:
a state $\rho_{AB}$ is separable iff its partial transpose $\rho_{AB}^{\mathrm{T}_{B}}$
with respect to the second subsystem is positive semi-definite. The
Peres-Horodecki criterion is also called Positive Partial Transpose
(PPT) criterion.


The \textit{concurrence} is an entanglement monotone, and it has a direct
relationship with \textit{entanglement of formation}~\citep{Hill:1997prl}.
In this work, we utilize the concurrence as a measure of the entanglement.
In Ref.~\citep{Wootters:1998prl}, Wootters derived the two-qubit
concurrence as
\begin{equation}
\mathcal{C}[\rho]\equiv\max\left\{ 0,\mu_{1}-\mu_{2}-\mu_{3}-\mu_{4}\right\} ,
\end{equation}
where $\mu_{i}$ with $i=1,2,3,4$ are the eigenvalues of the Hermitain
matrix $\sqrt{\sqrt{\rho}\tilde{\rho}\sqrt{\rho}}$ with $\tilde{\rho}=(\sigma_{y}\otimes\sigma_{y})\rho^{*}(\sigma_{y}\otimes\sigma_{y})$
in the decreasing order, and $\rho^{*}$ denotes the complex conjugate
of $\rho$ in the spin basis of $\sigma_{z}$. Wootters' concurrence
is a function in the range $[0,1]$. A state is separable for $\mathcal{C}[\rho]=0$
and is entangled for $\mathcal{C}[\rho]>0$. When $\mathcal{C}[\rho]=1$,
the state is said to be maximally entangled.


We rewrite the spin density operator for the hyperon-antihyperon system
in the $\sigma_{z}$ basis 
\begin{equation}
\rho_{Y\bar{Y}}^{X}=\frac{1}{4}\begin{bmatrix}1+2a+t_{3} & 0 & 0 & t_{1}-t_{2}\\
0 & 1-t_{3} & t_{1}+t_{2} & 0\\
0 & t_{1}+t_{2} & 1-t_{3} & 0\\
t_{1}-t_{2} & 0 & 0 & 1-2a+t_{3}
\end{bmatrix},\label{eq:X_form}
\end{equation}
where $a$ and $t_{1,2,3}$ are defined in Eq.~(\ref{eq:a_and_t}).
The above expression can be directly obtained by expanding Pauli operators
in Eq.~(\ref{eq:rho_X}) into a $2\times2$ matrix form. The name
$X$ state comes from its resemblance to the letter $X$.


The Peres-Horodecki criterion for a general $X$ state claims that
the state is entangled iff either $\rho_{22}^{X}\rho_{33}^{X}<|\rho_{14}^{X}|^{2}$
or $\rho_{11}^{X}\rho_{44}^{X}<|\rho_{23}^{X}|^{2}$ holds~\citep{Hiroo:2010jmo},
but both conditions cannot be satisfied simultaneously~\citep{Sanpera:1998pra}.
The Wootters' concurrence for the $X$ state is given by~\citep{Ting:2007qic}
\begin{equation}
\mathcal{C}\left[\rho^{X}\right]=2\max\left\{ 0,|\rho_{14}^{X}|-\sqrt{\rho_{22}^{X}\rho_{33}^{X}},|\rho_{23}^{X}|-\sqrt{\rho_{11}^{X}\rho_{44}^{X}}\right\} ,\label{eq:concurrence}
\end{equation}
with $\rho_{ij}^{X}$ being given in (\ref{eq:X_form}). We see that
the Peres-Horodecki criterion for the $X$ state is compatible with
the concurrence.


\begin{figure}[ht]
\includegraphics[scale=0.46]{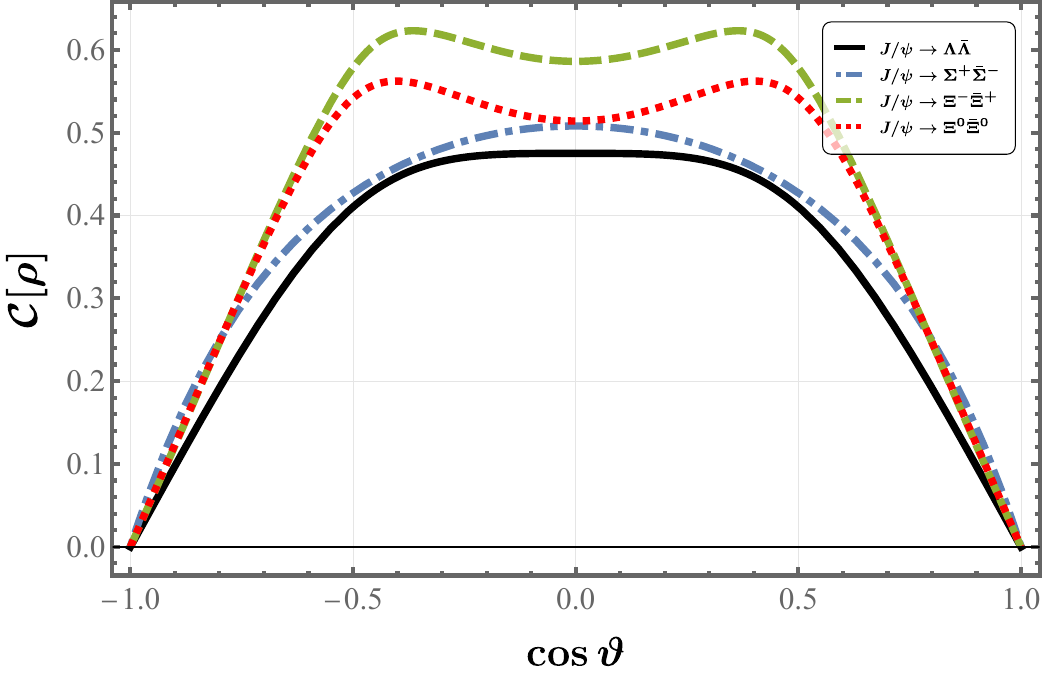}

\caption{\label{fig:concurrence}Wootters' concurrence $\mathcal{C}[\rho_{Y\bar{Y}}^{X}]$
as functions of $\cos\vartheta$ ($\vartheta$ is the scattering angle),
where $Y=\Lambda$, $\Sigma^{+}$, $\Xi^{-}$ and $\Xi^{0}$ corresponding
to curves in black solid, blue dash-dotted, green dashed, and red
dotted lines, respectively. The black horizontal line is the entanglement bound. The $Y\bar{Y}$ system is entangled iff $\mathcal{C}>0$.}
\end{figure}


With Eqs.~(\ref{eq:X_form}) and (\ref{eq:concurrence}), we derive
the concurrence for the hyperon-antihyperon system as 
\begin{align}
 & \mathcal{C}\left[\rho_{Y\bar{Y}}^{X}\right]=\left|t_{2}\right|\nonumber \\
 & =\frac{\left|1+\alpha_{\psi}-\sqrt{\left(1+\alpha_{\psi}\cos2\vartheta\right)^{2}-\beta_{\psi}^{2}\sin^{2}2\vartheta}\right|}{2(1+\alpha_{\psi}\cos^{2}\vartheta)}.\label{eq:concurrence_t}
\end{align}
The results for the concurrence as functions of $\vartheta$ for octet
hyperons listed in Table~\ref{tab:decay_parameters} are shown
in Fig.~\ref{fig:concurrence}. We see that the entanglement of $Y\bar{Y}$ pairs exists 
in the whole range of the scattering angle $\vartheta$ except at two collinear limits $\vartheta=0$ or $\pi$. However, unlike the Bell nonlocality, the maximum value of the concurrence (or maximum entanglement) does not necessarily take place at $\vartheta = \pi/2$. Instead, it can occur at other angles such as the ones for $\Xi^{-}\bar{\Xi}^{+}$ and $\Xi^{0}\bar{\Xi}^{0}$ shown in Table~\ref{tab:concurrence}. 

\begin{table}
\caption{\label{tab:concurrence}The maximum concurrence $\mathcal{C}_{\max}$ in Eq.~(\ref{eq:concurrence_t})
and their corresponding angles $\vartheta_{\max}$ in  $e^{+}e^{-}\rightarrow J/\psi\rightarrow Y\bar{Y}$.}
\begin{ruledtabular}
\begin{tabular}{ccccc}
 & $\Lambda\bar{\Lambda}$ & $\Sigma^{+}\bar{\Sigma}^{-}$ & $\Xi^{-}\bar{\Xi}^{+}$ & $\Xi^{0}\bar{\Xi}^{0}$\tabularnewline
\hline 
$\mathcal{C}_{\max}$ & 0.475 & 0.508 & 0.623 & 0.562
\tabularnewline
$\vartheta_{\max}$ & 90\textdegree & 90\textdegree & 68.60\textdegree, 111.40\textdegree & 66.26\textdegree, 113.74\textdegree 
\tabularnewline
\end{tabular}
\end{ruledtabular}

\end{table}


In summary, the outgoing hyperon-antihyperon pairs are entangled in
the full range of the scattering angle except at two boundaries.


\section{Discussions on Bell nonlocality and entanglement \label{sec:Discussions}}

In this section, we will discuss the relation between Bell nonlocality
and entanglement, the eigenvalue decomposition of the spin density
matrix, the role of electromagnetic form factors in quantum correlation
of the hyperon-antihyperon system. 


\subsection{Bell nonlocality versus entanglement}

Given that both Bell nonlocality and quantum entanglement characterize
quantum properties of a system, we try to look for the relationship
between them.

\begin{figure}[ht]
\begin{centering}
\includegraphics[scale=0.24]{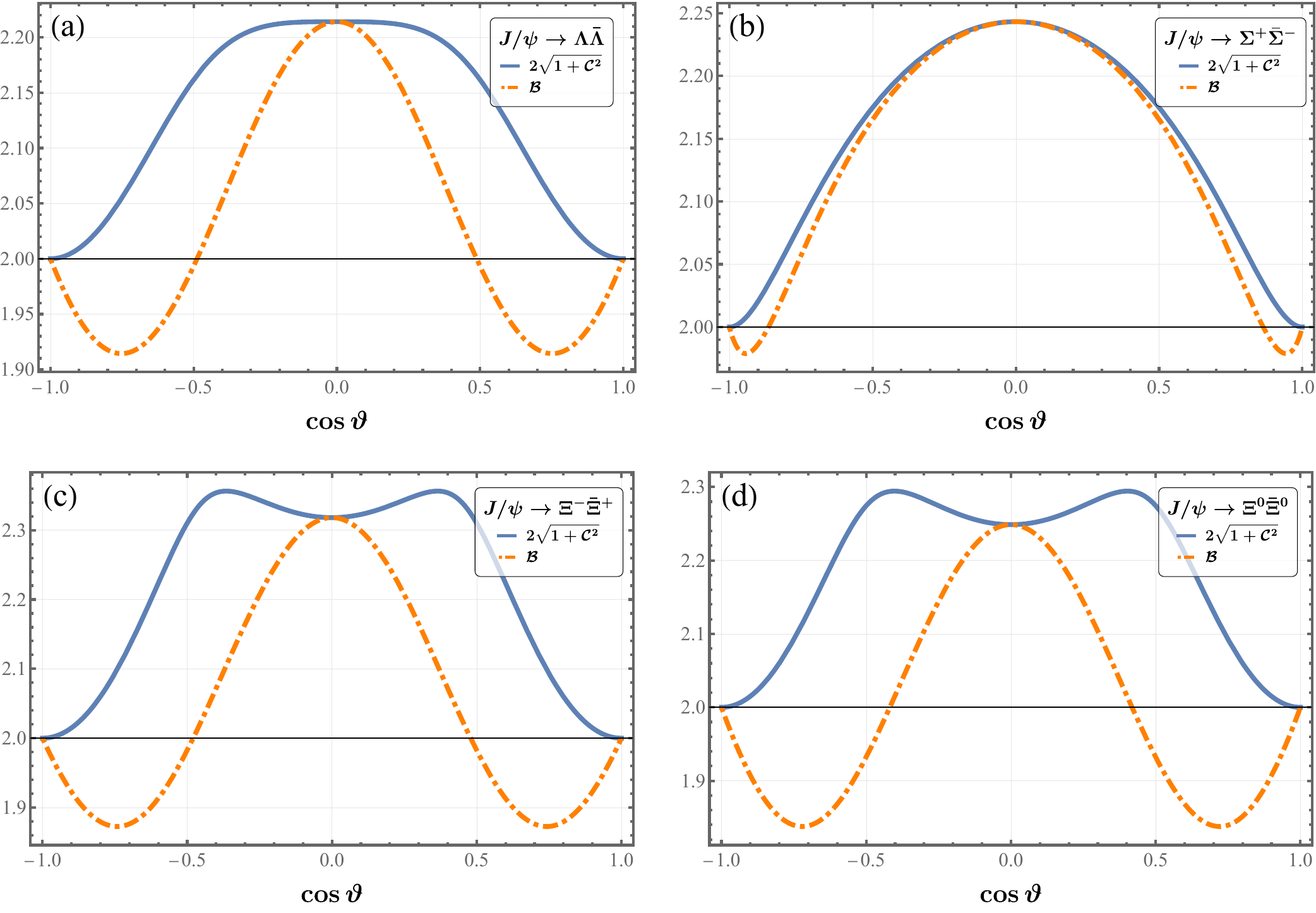}
\par\end{centering}
\caption{\label{fig:nonlocality_vs_entanglement}The measures $\mathcal{B}=2\sqrt{\mathfrak{m}_{12}}$
and $2\sqrt{1+\mathcal{C}^{2}}$ for the Bell nonlocality and quantum
entanglement as functions of $\cos\vartheta$ ($\vartheta$ is the
scattering angle). The four panels (a)-(d) correspond to four decay
channels of $J/\psi$ to $Y\bar{Y}$ with $Y=\Lambda$, $\Sigma^{+}$,
$\Xi^{-}$ and $\Xi^{0}$, respectively. Solid blue lines are curves
of $2\sqrt{1+\mathcal{C}^{2}}$ for the entanglement, while orange
dot-dashed lines are curves of $\mathcal{B}$. The black solid horizontal
line is the value 2. The $Y\bar{Y}$ system is nonlocal or entangled
iff $\mathcal{B}>2$ or $2\sqrt{1+\mathcal{C}^{2}}>2$. }
\end{figure}

For a two-qubit density operator $\rho$ with Wootters' concurrence
$\mathcal{C}[\rho]$, the maximum violation of the CHSH inequality
$\mathcal{B}[\rho]$ has an upper bound~\citep{Verstraete:2002prl}
\begin{equation}
\mathcal{B}[\rho]\leq2\sqrt{1+\mathcal{C}^{2}[\rho]},\label{eq:nonlocal_bound}
\end{equation}
with $\mathcal{B}[\rho]\equiv2\sqrt{\mathfrak{m}_{12}}$ defined in
Eq.~(\ref{eq:Bell_violation}). In Fig.~\ref{fig:nonlocality_vs_entanglement}
we plot $\mathcal{B}$ and $2\sqrt{1+\mathcal{C}^{2}}$ as functions
of $\cos\vartheta$. We see in Fig.~\ref{fig:nonlocality_vs_entanglement}
that the inequality (\ref{eq:nonlocal_bound}) is always satisfied
and the equality $\mathcal{B}=2\sqrt{1+\mathcal{C}^{2}}$ (or equivalently
$\mathfrak{m}_{12}=1+\mathcal{C}^{2}$) holds at $\vartheta=\pi/2$.
At this transverse scattering angle, $Y$'s and $\bar{Y}$'s polarizations
vanish from Eq.~(\ref{eq:py}), then the spin density operator $\rho_{Y\bar{Y}}^{X}$
reduces to a very special subclass of the $X$ state: \textit{$T$
state} or \textit{Bell Diagonal State~(BDS). }The upper bound of
$\mathcal{B}$ in (\ref{eq:nonlocal_bound}) is attained for rank-2
BDSs~\citep{Verstraete:2002prl}. 


From Fig.~\ref{fig:nonlocality_vs_entanglement}, both measures for
the Bell nonlocality and entanglement are symmetric with respect to
$\vartheta=\pi/2$. However, even if hyperon-antihyperon pairs are
entangled in the full range of scattering angle except at $\vartheta=0$ or $\pi$, the Bell nonlocality only appears in the range $\vartheta\in(\vartheta^{*},\pi-\vartheta^{*})$. This corresponds
to the range where orange dot-dashed lines lie above the black line
in Fig.~\ref{fig:nonlocality_vs_entanglement}. This indicates the relation between
the Bell nonlocality and entanglement in the hierarchy of quantumness
\begin{equation}
\textrm{Bell nonlocality}\subset\textrm{entanglement}. \nonumber
\end{equation}
Any nonlocal state must be entangled, but not
all entangled states can have nonlocal correlation~\citep{Werner:1989pra}.


Another interesting behavior of the entanglement and nonlocality appears in the panels (c) and (d) in Fig.~\ref{fig:nonlocality_vs_entanglement} for $\Xi^{-}\bar{\Xi}^{+}$ and $\Xi^{0}\bar{\Xi}^{0}$: the entanglement in the range from the maximum concurrence angle $\vartheta_{\max}$ 
(see Table~\ref{tab:concurrence}) to $\pi/2$ shows a decreasing trend 
while the Bell nonlocality is still increasing. This phenomenon, where less entanglement corresponds to more nonlocality, sometimes referred to as an \textit{anomaly of nonlocality}~\citep{Methot:2007qic}. It can be explained by the quantum resource theory that the entanglement and nonlocality may be inequivalent resources~\citep{Chitambar:2019rmp}. The subtle relationship between the entanglement and nonlocality 
is still an active topic in this field~\citep{Lipinska:2018njop}.


\subsection{Eigenvalue decomposition}

Any two-qubit density operator can be decomposed as $\rho=\sum_{i=1}^{4}\lambda_{i}|\lambda_{i}\rangle\langle\lambda_{i}|$,
with $\lambda_{i}$ being the eigenvalue and $|\lambda_{i}\rangle$
the corresponding eigenstate. According to Eq.~(\ref{eq:X_form}),
the spin density operator has only two non-zero eigenvalues 
\begin{equation}
\lambda_{1,2}=\frac{1}{2}\left(1\mp\frac{\alpha_{\psi}\sin^{2}\vartheta}{1+\alpha_{\psi}\cos^{2}\vartheta}\right),\label{eq:eigenvalues}
\end{equation}
for the corresponding eigenstates 
\begin{align}
\left|\lambda_{1}\right\rangle  & =\sqrt{\frac{1+\alpha_{\psi}\cos2\vartheta+\beta_{\psi}\sin2\vartheta}{2(1+\alpha_{\psi}\cos2\vartheta)}}\left|00\right\rangle \nonumber \\
 & \ +\sqrt{\frac{1+\alpha_{\psi}\cos2\vartheta-\beta_{\psi}\sin2\vartheta}{2(1+\alpha_{\psi}\cos2\vartheta)}}\left|11\right\rangle ,\nonumber \\
\left|\lambda_{2}\right\rangle  & =\frac{1}{\sqrt{2}}\left(\left|01\right\rangle +\left|10\right\rangle \right),\label{eq:eigenvectors}
\end{align}
where we adopt the notation for spin states: $|0\rangle\equiv|\uparrow_{z}\rangle$,
$|1\rangle\equiv|\downarrow_{z}\rangle$. Through the eigenvalue decomposition,
the spin configuration can be clearly shown in Eqs.~(\ref{eq:eigenvalues})
and (\ref{eq:eigenvectors}) that $\rho_{Y\bar{Y}}^{X}$ can be treated
as an ensemble of two pure states $\{|\lambda_{1}\rangle,|\lambda_{2}\rangle\}$
with probabilities $\{\lambda_{1},\lambda_{2}\}$.


The eigenstate $|\lambda_{1}\rangle$ is a superposition of two spin
triplet states: $|00\rangle=|S=1,S_{z}=1\rangle$ and $|11\rangle=|S=1,S_{z}=-1\rangle$,
and $|\lambda_{2}\rangle$ is another spin triplet state: $|S=1,S_{z}=0\rangle$.
We see that there is no spin singlet component in the $Y\bar{Y}$
system. This is the result of the angular momentum conversation in
$J/\psi$'s decay, and it coincides with the partial wave analysis
that the outgoing $Y\bar{Y}$ only has contributions from $^{3}S_{1}$
and $^{3}D_{1}$ waves~\citep{Haidenbauer:2016won}.


The lack of spin singlet component can also be seen by imposing the
spin projection operator $F_{S}=(1-\boldsymbol{\sigma}\cdot\boldsymbol{\sigma})/4$
on the spin density matrix~\citep{Barnes:1990bs} as 
\begin{equation}
\mathrm{Tr}\left\{ \rho_{Y\bar{Y}}^{X}F_{S}\right\} =\mathrm{Tr}\,\mathbf{t}=\mathrm{Tr}\,C=0,
\end{equation}
with $\mathbf{t}$ and $C$ being the correlation matrix in Eqs.~(\ref{eq:rho_X})
and (\ref{eq:4by4}) respectively.


\subsection{Electromagnetic form factors}

In this subsection, we will look into the time-like electromagnetic
form factors (EMFFs) in $e^{+}e^{-}\rightarrow Y\bar{Y}$ and investigate
their effects on nonlocality and entanglement.

The electromagnetic current of the spin-1/2 hyperon can be expressed
in terms of the Dirac form factor $F_{1}$ and Pauli form factor $F_{2}$
as~\citep{Faldt:2017kgy}
\begin{equation}
\Gamma^{\mu}=\gamma^{\mu}F_{1}(q^{2})+i\frac{\sigma^{\mu\nu}q_{\nu}}{2M}F_{2}(q^{2}),\label{eq:vertex}
\end{equation}
where $q=p_{1}+p_{2}$ with $p_{1}$ and $p_{2}$ being the four-momentum
of the hyperon and antihyperon respectively, and $M$ is the hyperon
mass. With $s=q^{2}$, the electric and magnetic form factors $G_{E}$
and $G_{M}$ are related to $F_{1}$ and $F_{2}$ by 
\begin{equation}
G_{E}(s)=F_{1}+\frac{s}{4M^{2}}F_{2},\ \ G_{M}(s)=F_{1}+F_{2}.
\end{equation}
Two parameters $\alpha_{\psi}$ and $\Delta\Phi$ in the process $e^{+}e^{-}\rightarrow Y\bar{Y}$
are related to $G_{E}$ and $G_{M}$ by 
\begin{eqnarray}
\alpha_{\psi} & = & \frac{s-4M^{2}\left|G_{E}/G_{M}\right|^{2}}{s+4M^{2}\left|G_{E}/G_{M}\right|^{2}}\in[-1,1],\nonumber \\
\Delta\Phi & = & \arg\left\{ G_{E}/G_{M}\right\} \in(-\pi,\pi].
\label{eq:phase-ge-gm}
\end{eqnarray}


From Eq.~(\ref{eq:py}), nonvanishing polarizations of $Y$ and $\bar{Y}$
produced in annihilation of unpolarized electron and positron require
$\Delta\Phi\neq0$ and $\pi$. At the limit $\Delta\Phi=0$ or $\pi$,
however, there is only the spin correlation part in $\rho_{Y\bar{Y}}^{X}$
and without polarizations part from Eq.~(\ref{eq:rho_X}). This indicates
that $\rho_{Y\bar{Y}}$ is reduced to a BDS form as 
\begin{equation}
\rho_{Y\bar{Y}}^{\mathrm{BDS}}=\frac{1}{4}\biggl(\mathds{1}\otimes\mathds{1}+\sum_{i}t_{i}\sigma_{i}\otimes\sigma_{i}\biggr),
\end{equation}
where $t_{2}^{2}=t_{3}^{2}$. We note that a BDS is also a $X$ state
but without polarization.


Following Eqs.~(\ref{eq:m_12}) and (\ref{eq:t_12}), the measure of the
Bell nonlocality becomes 
\begin{equation}
\mathfrak{m}_{12}\left[\rho_{Y\bar{Y}}^{\mathrm{BDS}}\right]=1+\left(\frac{\alpha_{\psi}\sin^{2}\vartheta}{1+\alpha_{\psi}\cos^{2}\vartheta}\right)^{2}\geq1.\label{eq:nonlocality_BDS}
\end{equation}
We see in this circumstance the violation of the CHSH inequality occurs
in the full range of the scattering angle $\vartheta\in(0,\pi)$ for
any $\alpha_{\psi}\neq0$. This result is different from what we discussed
in Sec.~\ref{sec:Nonlocality}, where the Bell nonlocality is violated
in a restricted angle range $(\vartheta^{*},\pi-\vartheta^{*})$.
However, the maximal violation of the CHSH inequality also takes place
at $\vartheta=\pi/2$ with the value in (\ref{eq:max_violation}).


The concurrence in Eq.~(\ref{eq:concurrence_t}) for a BDS is reduced
to 
\begin{equation}
\mathcal{C}\left[\rho_{Y\bar{Y}}^{\mathrm{BDS}}\right]=\frac{|\alpha_{\psi}|\sin^{2}\vartheta}{1+\alpha_{\psi}\cos^{2}\vartheta}.\label{eq:concurrence_BDS}
\end{equation}
Comparing Eq.~(\ref{eq:nonlocality_BDS}) and (\ref{eq:concurrence_BDS}),
one can see that the inequality in Eq.~(\ref{eq:nonlocal_bound})
becomes an equality $\mathcal{B}=2\sqrt{1+\mathcal{C}^{2}}$ (or equivalently
$\mathfrak{m}_{12}=1+\mathcal{C}^{2}$) in the whole range of the
scattering angle (not only at $\vartheta=\pi/2$). It is not a surprise
since the property $\mathcal{B}=2\sqrt{1+\mathcal{C}^{2}}$ (or equivalently
$\mathfrak{m}_{12}=1+\mathcal{C}^{2}$) is valid for any rank-$2$
BDS~\citep{Verstraete:2002prl} with the fact that both $|\lambda_{1}\rangle$
and $|\lambda_{2}\rangle$ become two Bell states $(|00\rangle+|11\rangle)/\sqrt{2}$
and $(|01\rangle+|10\rangle)/\sqrt{2}$ with $\beta_{\psi}=0$.


From Eq.~(\ref{eq:phase-ge-gm}) we see that $\Delta\Phi$ is the relative
phase between $G_{E}$ and $G_{M}$. Let us take an example for the
limit case $\Delta\Phi=0,\pi$ by assuming $G_{E}=\pm G_{M}$. As
a consequence, the measures for the nonlocality and Wootters' concurrence
are given by 
\begin{eqnarray}
\mathfrak{m}_{12} & = & 1+\left[\frac{\left(s-4M^{2}\right)\sin^{2}\vartheta}{4M^{2}\sin^{2}\vartheta+s\left(\cos^{2}\vartheta+1\right)}\right]^{2},\nonumber \\
\mathcal{C} & = & \frac{\left(s-4M^{2}\right)\sin^{2}\vartheta}{4M^{2}\sin^{2}\vartheta+s\left(\cos^{2}\vartheta+1\right)}.
\end{eqnarray}
The above expressions coincide with Eqs. (3.4) and (3.7) in Ref.~\citep{Ehataht:2023zzt}
for $e^{+}e^{-}\rightarrow\tau^{+}\tau^{-}$. This is reasonable since
the vertex Eq.~(\ref{eq:vertex}) in $e^{+}e^{-}\rightarrow\tau^{+}\tau^{-}$
is simply $\gamma^{\mu}$ indicating $G_{E}=G_{M}$.


In the process $e^{+}e^{-}\rightarrow Y\bar{Y}$, the existence of
EMFFs manifests in a polarized final state, even if the colliding
beams are unpolarized~\citep{Dubnickova:1996nca}. And this polarization
effect leads to the $Y\bar{Y}$ spin correlation different from that
in processes $e^{+}e^{-}\rightarrow\tau^{+}\tau^{-}$ and $pp\rightarrow t\bar{t}$
pairs~\citep{Ehataht:2023zzt,Afik:2020onf,Afik:2022kwm}.


\section{Quantum tomography in experiments \label{sec:experiment}}

The spin polarization of the hyperon and antihyperon can be measured
through their weak decays \citep{Cronin:1963zb,Pondrom:1985aw,BESIII:2021ypr}
$Y\rightarrow BM$ and $\bar{Y}\rightarrow\bar{B}\bar{M}$. The spin
correlation in $Y\bar{Y}$ can also be extracted from the joint decay
$Y\bar{Y}\rightarrow B\bar{B}(M\bar{M})$ through the joint angular
distribution of $B\bar{B}$~\citep{Wu:2024mtj}
\begin{align}
I(\vartheta;\theta,\bar{\theta})= & \frac{1}{(4\pi)^{2}}\biggl[1+\alpha_{Y}\sum_{i}P_{i}^{+}(\vartheta)\cos\theta_{i}\nonumber \\
 & +\alpha_{\bar{Y}}\sum_{j}P_{j}^{-}(\vartheta)\cos\bar{\theta}_{j}\nonumber \\
 & +\alpha_{Y}\alpha_{\bar{Y}}\sum_{i,j}C_{ij}(\vartheta)\cos\theta_{i}\cos\bar{\theta}_{j}\biggr],\label{eq:joint_angular}
\end{align}
where $i,j=1,2,3$ or $x,y,z$ denote three directions in the rest
frame of $Y$ and $\bar{Y}$ respectively, $\cos\theta_{i}$ and $\cos\bar{\theta}_{j}$
are projections of $B$ and $\bar{B}$'s momentum directions onto
the axis $i$ and $j$ respectively, and $\alpha_{Y}$ and $\alpha_{\bar{Y}}$
are the decay parameters in $Y\rightarrow BM$ and $\bar{Y}\rightarrow\bar{B}\bar{M}$
respectively.


\begin{table}[h]
\caption{\label{tab:weak_decay}Decay parameters for ground-state octet hyperons.
In our analysis, we neglect the $\mathcal{CP}$ violation effect so
we have $\alpha_{\bar{Y}}=-\alpha_{Y}$.}

\begin{ruledtabular}
\begin{tabular}{cccc}
$Y$ & $\mathscr{B}(\%)$ & $\alpha_{Y}$ & Ref\tabularnewline
\hline 
$\Lambda\rightarrow p\pi^{-}$ & $064$ & $0.755(3)$ & \citep{BESIII:2021ypr,BESIII:2022qax}\tabularnewline
$\Sigma^{+}\rightarrow p\pi^{0}$ & $052$ & $-0.994(4)$ & \citep{BESIII:2020fqg}\tabularnewline
$\Xi^{-}\rightarrow\Lambda\pi^{-}$ & $100$ & $-0.379(4)$ & \citep{BESIII:2021ypr,ParticleDataGroup:2022pth}\tabularnewline
$\Xi^{0}\rightarrow\Lambda\pi^{0}$ & $96$ & $-0.375(3)$ & \citep{ParticleDataGroup:2022pth,BESIII:2023drj}\tabularnewline
\end{tabular}
\end{ruledtabular}

\end{table}

By adopting the idea of the quantum tomography~\citep{Afik:2020onf,Bernal:2023jba}
and the method of moments, the spin polarization and correlation in
the hyperon-antihyperon system can be extracted from the joint distribution
(\ref{eq:joint_angular}) as 
\begin{align}
P_{i}^{+}(\vartheta) & =\frac{3}{\alpha_{Y}}\left\langle \cos\theta_{i}\right\rangle ,\ P_{j}^{-}=\frac{3}{\alpha_{\bar{Y}}}\left\langle \cos\bar{\theta}_{j}\right\rangle ,\nonumber \\
C_{ij}(\vartheta) & =\frac{9}{\alpha_{Y}\alpha_{\bar{Y}}}\left\langle \cos\theta_{i}\cos\bar{\theta}_{j}\right\rangle .
\end{align}
In this way, $15$ real parameters $\mathbf{P}^{\pm}$ and $C_{ij}$
in $\rho_{Y\bar{Y}}$ in Eq.~(\ref{eq:two_qubit}) can be constructed
from experiment data.


Furthermore, due to parity and charge conjugation invariance, these
15 parameters are not all independent: the only non-zero polarization
is perpendicular to the production plane (i.e. in $\hat{\mathbf{y}}$
direction) and $P_{y}^{+}=P_{y}^{-}$. The correlation is a symmetric
matrix $C_{ij}=C_{ji}$ with $C_{xy}=C_{yz}=0$. Then the $4\times4$
matrix $\Theta_{\mu\nu}$ reads 
\begin{equation}
\Theta_{\mu\nu}(\vartheta)=\left[\begin{array}{c|ccc}
1 & 0 & P_{y} & 0\\
\hline 0 & C_{xx} & 0 & C_{xz}\\
P_{y} & 0 & C_{yy} & 0\\
0 & C_{xz} & 0 & C_{zz}
\end{array}\right],\label{eq:tomo_Theta}
\end{equation}
where all elements are functions of the scattering angle $\vartheta$.
Obviously, Eq.~(\ref{eq:tomo_Theta}) is local unitary equivalent
to the standard $X$ state.


The Bell nonlocality measure $\mathfrak{m}_{12}$ is given by the
sum of two largest eigenvalues $C^{\mathrm{T}}C$ whose three eigenvalues
of $C^{\mathrm{T}}C$ are 
\begin{equation}
C_{yy}^{2},\ \ \frac{1}{4}\left[C_{xx}+C_{zz}\pm\sqrt{4C_{xz}^{2}+\left(C_{xx}-C_{zz}\right)^{2}}\right]^{2}.
\end{equation}
The concurrence $\mathcal{C}$ is given by 
\begin{align}
\mathcal{C}=\frac{1}{2}\max\biggl\{ & 0,\sqrt{4C_{xz}^{2}+\left(C_{xx}-C_{zz}\right)^{2}}-\left|1-C_{yy}\right|,\nonumber \\
 & \left|C_{xx}+C_{zz}\right|-\sqrt{\left(1+C_{yy}\right)^{2}-4P_{y}^{2}}\biggr\}.
\end{align}
Since $P_{y}$ and $C_{xx}$, $C_{yy}$, $C_{zz}$ and $C_{xz}$ can
all be constructed from data, the Bell nonlocality and entanglement
can be tested in experiments.


The above probe to quantum correlation in $e^{+}e^{-}$ annihilation
at BESIII can also be extended to $p\bar{p}\rightarrow Y\bar{Y}$
at PANDA~\citep{PANDA:2020zwv}, in which the spin-parity of the
intermediate resonance is not necessarily $1^{-}$.


\section{Summary and outlook \label{sec:Summary}}

In this work, we present the study of the Bell nonlocality and entanglement
in $e^{+}e^{-}\rightarrow Y\bar{Y}$, with $Y$ being the spin-1/2
octet hyperon. We begin with the spin density operator for $Y\bar{Y}$
and convert it into that for the standard two-qubit $X$ state. Using
properties of $X$ states, we derive analytical formulas for the Bell
nonlocality and entanglement in various $Y\bar{Y}$ systems, based
on two intrinsic parameters, $\alpha_{\psi}$ and $\Delta\Phi$, along
with a kinematic variable, the scattering angle $\vartheta$. We explore
the relation between the Bell nonlocality and entanglement and present
the experimental proposal to test the nonlocality and entanglement
at BESIII.


In $e^{+}e^{-}\rightarrow Y\bar{Y}$, the relative phase between the
electric and magnetic form factors of hyperons lead to their polarizations
in the spin density operator. With nonvanishing polarizations of $Y$
and $\bar{Y}$, the kinematic region of nonlocality in the $Y\bar{Y}$
system is more restricted than $\tau^{+}\tau^{-}$~\citep{Ehataht:2023zzt,Fabbrichesi:2022ovb}
and and $t\bar{t}$ systems~\citep{Fabbrichesi:2021npl,Afik:2020onf,Afik:2022dgh,Afik:2022kwm}
where polarizations of tau leptons and top quarks are vanishing. The
entanglement in the $Y\bar{Y}$ system can also influenced by the
polarization effect in comparison with $\tau^{+}\tau^{-}$ and and
$t\bar{t}$ systems. This is the main result of our work.


Our work offers a theoretical framework for probing the nonlocality
and entanglement in hyperon-antihyperon systems at BESIII. Our method
can also be applied to other collision processes with $X$-form final
states such as $p\bar{p}\rightarrow Y\bar{Y}$ at PANDA~\citep{PANDA:2020zwv}.
A modified CHSH inequality and related entanglement measures were
proposed to quantify the quantum entanglement and spin correlation
of $\Lambda\bar{\Lambda}$ in string fragmentation \citep{Gong:2021bcp}.
Our method can also be generalized to describe the nonlocality and
entanglement of such hyperon-antihyperon systems in many-body states. 

Note added in the second version of this paper: 
we notice that a new paper appeared in the arxiv addressing similar problems 
but in a different angle~\cite{Fabbrichesi:2024rec}. 

\begin{acknowledgments}
This work is supported by the National Natural Science Foundation
of China (NSFC) under Grant Nos. 12135011 and 12305010.

\end{acknowledgments}

\bibliographystyle{apsrev4-2}
\bibliography{refs}

\end{document}